\documentclass[fleqn,10pt]{wlscirep}
\usepackage[utf8]{inputenc}
\usepackage[T1]{fontenc}

\usepackage{microtype}
\usepackage{graphicx}
\usepackage{subfigure}
\usepackage{hyperref}
\usepackage{booktabs} 

\title{Segregation Dynamics with Reinforcement Learning and Agent Based Modeling}

\usepackage{authblk}
\makeatletter
\renewcommand\AB@affilsepx{. \protect\Affilfont}
\makeatother
\author[1,2]{Egemen Sert}
\author[1]{Yaneer~Bar-Yam}
\author[1,3,*]{Alfredo J. Morales}

\affil[1]{New England Complex Systems Institute, Cambridge, MA}
\affil[2]{Department  of  Electrical  and  Electronics  Engineering,  Middle  East  Technical  University,  Ankara,  Turkey}
\affil[3]{MIT Media Lab, Cambridge, MA.}
\affil[*]{Corresponding author: alfredo@necsi.edu}

\date{}

\begin{abstract}
Societies are complex. Properties of social systems can be explained by the interplay and weaving of individual actions. Incentives are key to understand people's choices and decisions. For instance, individual preferences of where to live may lead to the emergence of social segregation. In this paper, we combine Reinforcement Learning (RL) with Agent Based Models (ABM) in order to address the self-organizing dynamics of social segregation and explore the space of possibilities that emerge from considering different types of incentives. Our model promotes the creation of interdependencies and interactions among multiple agents of two different kinds that want to segregate from each other. For this purpose, agents use Deep Q-Networks to make decisions based on the rules of the Schelling Segregation model and the Predator-Prey model.
Despite the segregation incentive, our experiments show that spatial integration can be achieved by establishing interdependencies among agents of different kinds. They also reveal that segregated areas are more probable to host older people than diverse areas, which attract younger ones. Through this work, we show that the combination of RL and ABMs can create an artificial environment for policy makers to observe potential and existing behaviors associated to incentives.
\end{abstract}

\begin{document}

\maketitle



\section{Introduction}
\label{introduction}

The recent availability of large datasets collected from various resources, such as digital transactions, location data and government census, is transforming the ways we study and understand social systems \cite{lazer2009computational}. Researchers and policy makers are able to observe and model social interactions and dynamics in great detail, including the structure of friendship networks \cite{eagle2009inferring}, the behavior of cities \cite{doi:10.1098/rsif.2016.1048}, politically polarized societies \cite{morales2015measuring}, or the spread of information on social media \cite{vosoughi2018spread}. These studies show the behaviors present in the data but do not explore the space of possibilities that human dynamics may evolve to. Robust policies should consider mechanisms to respond to every type of events \cite{ashby1991requisite}, including those that are very rare \cite{taleb2007black}. Therefore it is crucial to develop simulation environments such that potentially unobserved social dynamics can be assessed empirically.

Agent Based Modeling (ABM) is a generative approach to study social phenomena based on the interaction of individuals \cite{sayama2015introduction}. These models show how different types of individual behavior give rise to emergent macroscopic regularities \cite{schelling1971dynamic}, such as unequal wealth distributions \cite{epstein1996growing}, new political actors \cite{axelrod2006model}, multipolarity in interstate systems \cite{cederman1997emergent} and cultural differentiation \cite{axelrod1997dissemination}. Moreover, ABM allows testing core sociological theories against simulations \cite{epstein1996growing} with emphasis on heterogeneous, autonomous actors with bounded, spatial information \cite{epstein1999agent}. However, the rules of agent interactions are generally fixed which limits the exploration of the space of possible behaviors. 

Reinforcement Learning (RL) is a simulation method where agents become intelligent and create new, optimal behaviors based on the state of their environment and a previously defined structure of incentives. This method is referred as Multi-Agent Reinforcement Learning (MARL) if multiple agents are employed. 
Recently, the combination of RL with Deep Learning architectures achieve human level performance in complex tasks, including video gaming \cite{mnih2015human}, motion in harsh environments \cite{heess2017emergence}, and effective communication networks without assumptions \cite{sert2018optimizing}. Moreover, it has been recently applied to study societal dilemma and game theory problems \cite{lanctot2017unified} such as the emergence of cooperation \cite{de2006learning,leibo2017multi}, the Prisoner's Dilemma \cite{sandholm1996multiagent} and payoff matrices in equilibrium \cite{wunder2010classes}. Although Deep RL algorithms applied to multiple agents (MARL) can shed light on social phenomena, to the best of our knowledge, the applications of these methods has been confined to classical game-theoretic problems  \cite{zawadzki2014empirically} and drawing connections to real-world examples remains unexplored.

In this paper we extend the standard ABM of social segregation using MARL in order to explore the space of possible behaviors as we modify the structure of incentives and promote the interaction among agents of different kinds. The idea is to observe the behavior of agents that want to segregate from each other when interactions are promoted. We achieve the segregation dynamics by considering the rules from the Schelling model \cite{schelling1971dynamic}. The creation of interdependencies among agents of different kinds is inspired by the dynamics of the Predator-Prey model \cite{sayama2015introduction} where agents hunt each other. Our experiments show that spatial segregation diminishes as more interdependencies among agents of different kinds are added. Moreover, our results shed light on previously unknown behaviors regarding segregation and the age of individuals which we confirmed using Census data. These methods can be extended to study other type of social phenomena and inform policy makers on possible actions.

The organization of the paper is as follows: In Section \ref{setup} we explain the experimental setup. Section \ref{experiments} illustrates the experiment outcomes. In Section \ref{discussion} we conclude and discuss our results. Future improvements are presented in Section \ref{sup:future} in the Supplement.

\section{Methods}
\label{setup}

\begin{figure}
    \centering
    \includegraphics[width=0.6\textwidth]{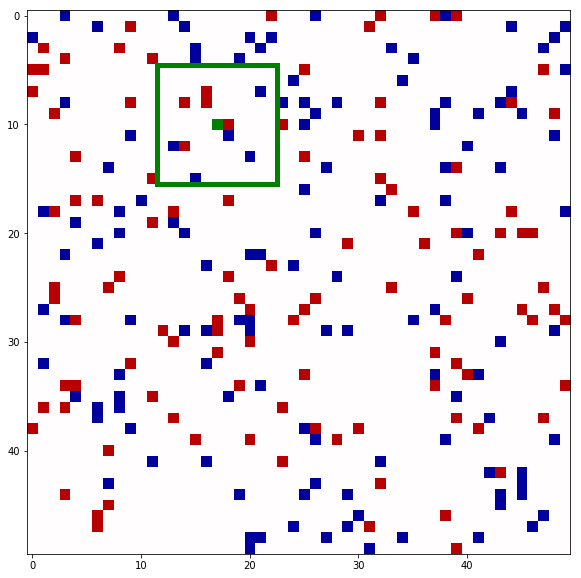} \\
\includegraphics[width=\textwidth]{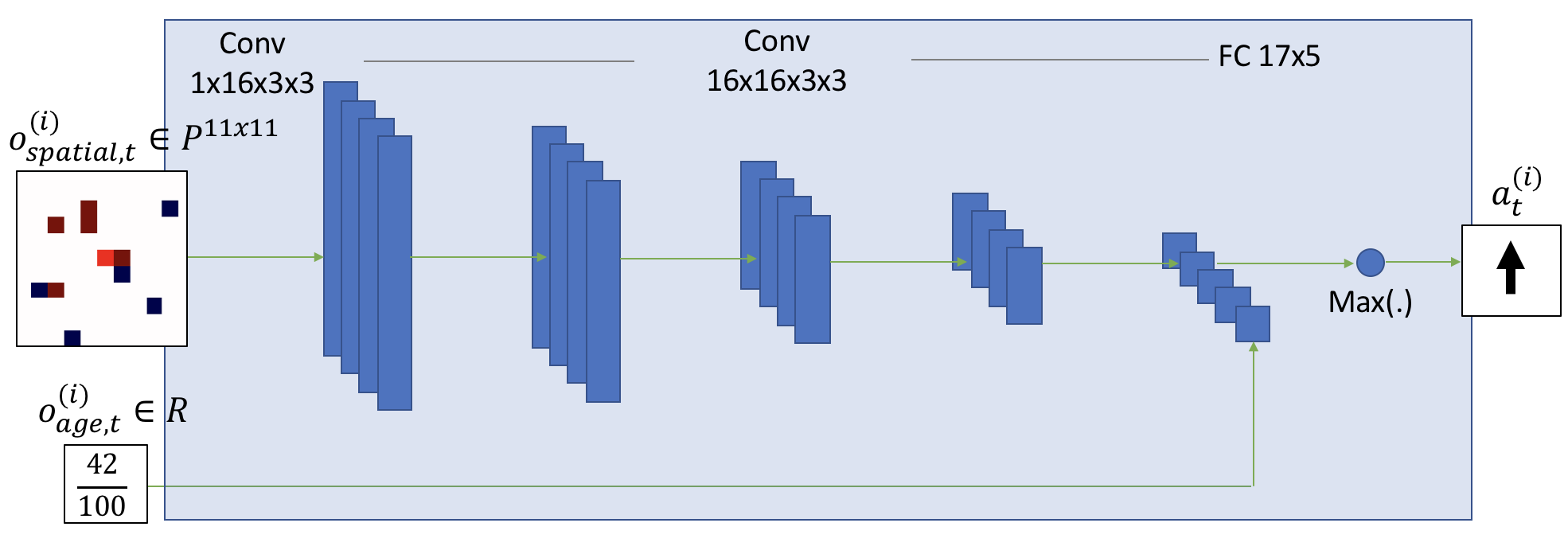}
    \caption{Schematic of the model simulation and network architecture. Top panel: Grid world of experiments. The grid size is 50x50 locations. Red and blue squares denote two types of agents respectively. White represents empty regions. Each type has its own Deep Q-Network. Every agent has a field of view of 11x11 locations. Green border denotes the field of view of the agent illustrated in green. Agents can move across empty spaces. Bottom panel: Network structure of $\phi_A$ and $\phi_B$. Each network receives an input of 11x11 locations, runs it through five convolution steps and concatenates the resulting activations with the agent's remaining age normalized by the maximum initial age. The feature vector is mapped over the action space using a fully connected layer. The action with the maximum Q-value is taken for the agent.}
    \label{fig:grid}
\end{figure}

\begin{table}[t]
\caption{Training parameters of the Deep Q-Networks used during the experiments.}
\label{tab:det}
\label{sample-table}
\vskip 0.15in
\begin{center}
\begin{small}
\begin{sc}
\begin{tabular}{lcccr}
\toprule
Parameter & Value & \\
\midrule
Number of Episodes                   & 1     \\ 
Batch Size                           & 256     \\
Number of Iterations                 & 5000     \\
Number of Training Steps             & 60.000     \\ 
Experience Memory Length             & 1.000.000     \\ 
Discount Factor ($\gamma$)                      & 0.98    \\
Learning Rate                        & 0.001   \\
Momentum                             & 0.999   \\
Double Network Copy Parameter ($\tau$) & 0.05    \\
Initial Exploration Rate             & 0.999   \\ 
Final Exploration Rate               & 0       \\ 
Exploration Decay (per agent action) & 100.000 \\ 
\bottomrule
\end{tabular}
\end{sc}
\end{small}
\end{center}
\vskip -0.1in
\end{table}

 We design a game in which agents are promoted to both self-segregate and interact with others. By varying the reward of interactions we are able to explore different incentives that affect the self-organizing process of segregation. Our experiments are based on two types of agents: A and B. Agents try to survive in a 50x50 grid where they can move around and interact with other agents. They observe an 11x11 patch of the grid centered around their current position and can live for a total of 100 iterations in isolation. Figure \ref{fig:grid} shows an schematic view of the grid world and the agents. Distinct colors indicate the agents' types and the green square represents the observation window of the agent illustrated in green.

\begin{figure*}
  \centering
  \includegraphics[width=0.8\linewidth]{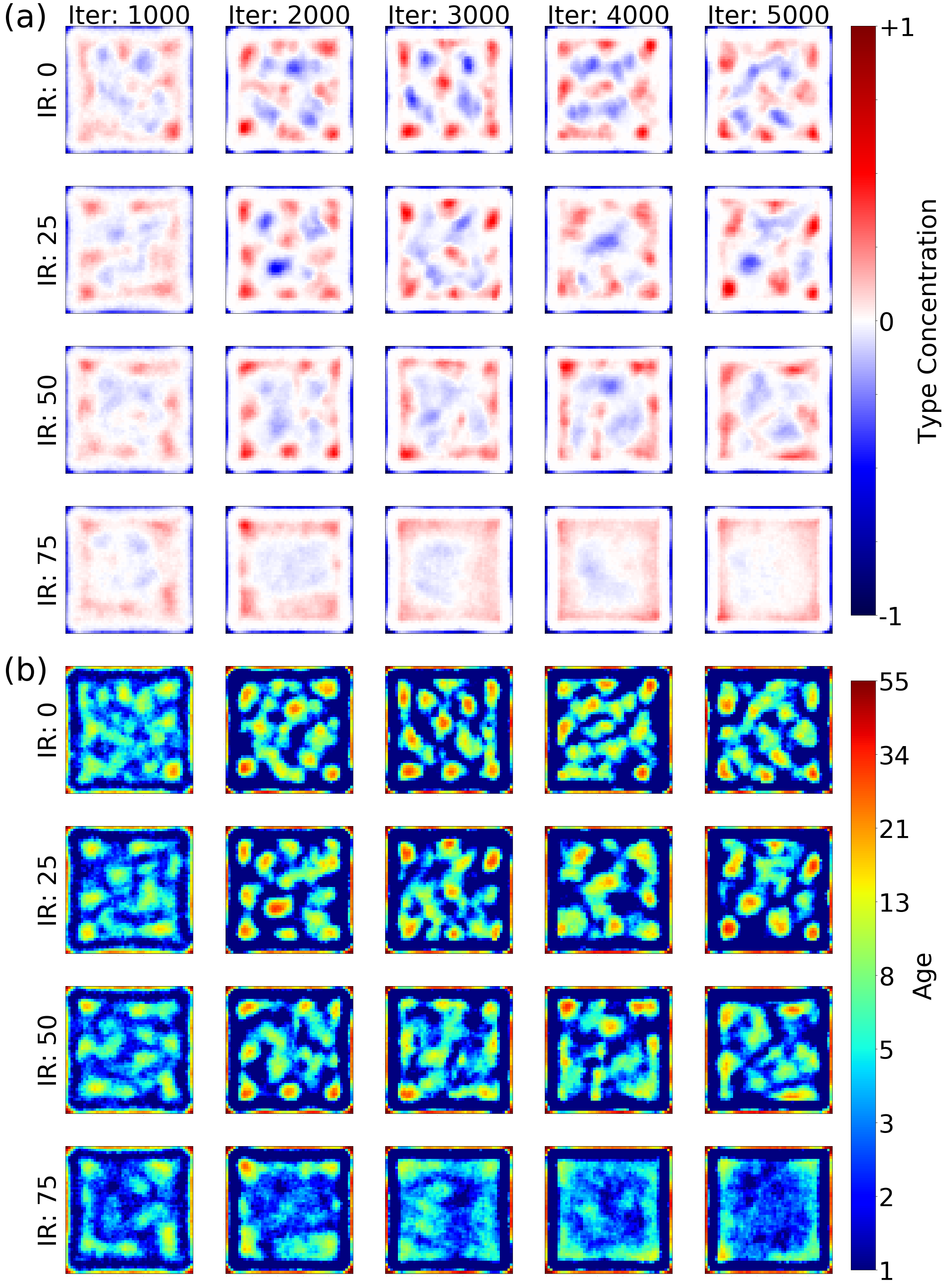}
  \caption{Agents collective behavior for multiple values of interdependence reward (rows) at multiple times (columns). Rows represent outcomes associated to different values of the interdependence reward (IR). Columns show the state of the system at different points of the simulation. Experiments are initialized with equal initial conditions and random seed. The heat maps are obtained by averaging over the last 1000 iterations. In Panel (a) red regions denote biased occupation of type A agents where areas fully occupied with type A agents are indicated by type concentration of +1. Blue regions denote biased occupation of type B agents where full occupation of blues are indicated by type concentration of -1. White areas indicate uniform mixing across types, indicated by type concentration of 0. In Panel (b) color indicates the age of agents irrespective of their type. As color shades from blue to red agent age increases. In Section \ref{experiments}, we introduce a set of videos that represent the experiments used in creating the heat maps.
  }
  \label{fig:heat}
\end{figure*}

\begin{figure}
    \centering
\includegraphics[width=0.8\textwidth]{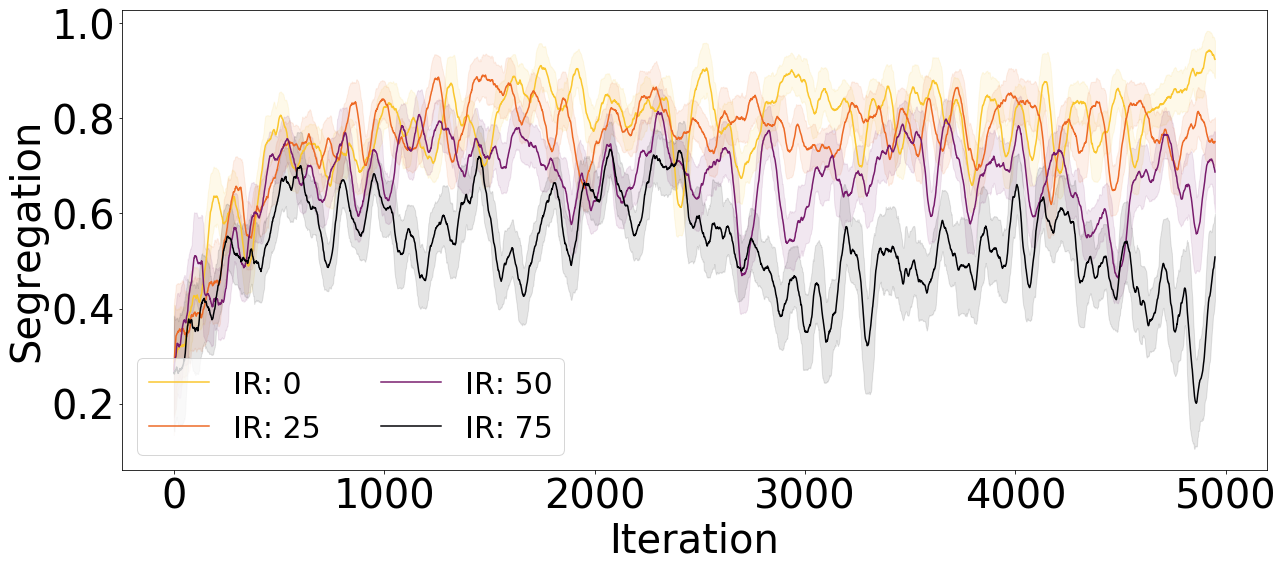}
    \caption{Segregation dynamics for multiple values of interdependence reward (IR). Colors correspond to the results for multiple values of interdependence reward, ranging from yellow (low) to black (high). The curves are obtained by averaging 50 iterations over 10 experiment realizations. Shades denote the standard deviation across experiments. } 
    \label{fig:segregation}
\end{figure}

Each type of agent utilizes one Deep Q-Network for maximizing rewards \cite{mnih2015human}. The rewards of the game, $R$, are as following:
\begin{itemize}
    \item \textbf{Segregation reward}. This incentive promotes agents to self-segregate. An agent is rewarded +1 for each agent of similar kind that joins its observation window, and -1 for each agent of different kind.
    \item \textbf{Interdependence reward}. This incentive promotes interactions among agents of different kinds. When an agent meets another agent of different kind, we randomly choose a winner of the interaction (following hunting dynamics). The winner (hunter) receives a positive reward, that we vary across experiments, and an extension of its lifetime by one iteration. 
    \item \textbf{Vigilance reward}. This incentive promotes agents to stay alive by providing +0.1 reward for every time step they survive.
    \item \textbf{Death reward}. This incentive rewards negatively agents who die or are hunted by agents of opposite kind. Agents receive -1 reward when they die.
    \item \textbf{Occlusion reward}. This incentive rewards movements towards occupied cells negatively. If an agent tries to move towards an occluded area, the agent receives -1 reward.
    \item \textbf{Stillness reward}. This incentive promotes the exploration of space. Agents who choose to stay still receive -1 reward.
\end{itemize}

Every agent takes one action at each iteration. The sequence of agents who take actions is chosen randomly. There are five possible actions for agents: to stay still or to move left, right, up or down. Agents are confined to the borders of the grid and cannot move towards agents of their own kind. If an agent moves to a location occupied by an agent of the opposite kind, it receives the interdependence reward and the opponent receives the death reward. 

Mathematically, agents of type A are represented as $-1$, B as $+1$, empty space as $0$ and border as $-2$ on the grid. Hence every agent's spatial observation at time $t$ is $O_{spatial, t}^{(i)} \in P^{11x11} ~ | ~ P \in \{-2, -1, 0, 1\}$. Moreover, every agent has the information of its remaining normalized life time, represented as $O_{age, t}^{(i)} \in R$. Full observation of the agent $i$ at time $t$ is $o_t^{i} \in O_t^{(i)} = O_{spatial,t}^{(i)} \cup O_{age,t}^{(i)}$. Let $\phi_A$ and $\phi_B$ denote the Q-Networks of type A and B. Then the networks' goal is to satisfy Equations \ref{eq:A} and \ref{eq:B}.

\begin{equation}
\phi_A^* = \arg\max_{\phi_A} \mathbb{E}[\sum_{t=0}^{T}\sum_{i=1}^{N_A} \gamma^{t}r_t^{(i)} |  o_t^{(i)}] \label{eq:A}
\end{equation}
\begin{equation}
\phi_B^* = \arg\max_{\phi_B} \mathbb{E}[\sum_{t=0}^{T}\sum_{i=1}^{N_B} \gamma^{t}r_t^{(i)} |  o_t^{(i)}] \label{eq:B} 
\end{equation}

where $N_X$ denotes the number of agents of type $X$, $\gamma$ denotes the discount factor, $r_t$ denotes the reward at time $t$ and $Q_{\phi_{X}}(.)$ denotes the Q-Network of agents of type $X$. 

Each network is initialized with the same parameters. In order to homogenize the networks' inputs, we normalize the observation windows by the agents' own kind, such that positive and negative values respectively represent equal and opposite kind for each agent. Actions are taken by following $\epsilon$-Greedy exploration strategy. Exploration rate decays exponentially. In order to stabilize the learning process, we use Adam optimizer \cite{kingma2014adam}, Experience Replay \cite{lin1992self} and Double Q-Learning \cite{van2016deep}. Networks are trained in parallel over 12 CPUs using data parallelism. We run one episode per experiment. Each episode is comprised of 5000 iterations. Each experiment is repeated 10 times for statistical analysis. Network details are given in Figure \ref{fig:grid} (bottom) and training details are given in Table \ref{tab:det}.



\section{Results}
\label{experiments}

\begin{figure}
    \centering
    \includegraphics[width=0.6\textwidth]{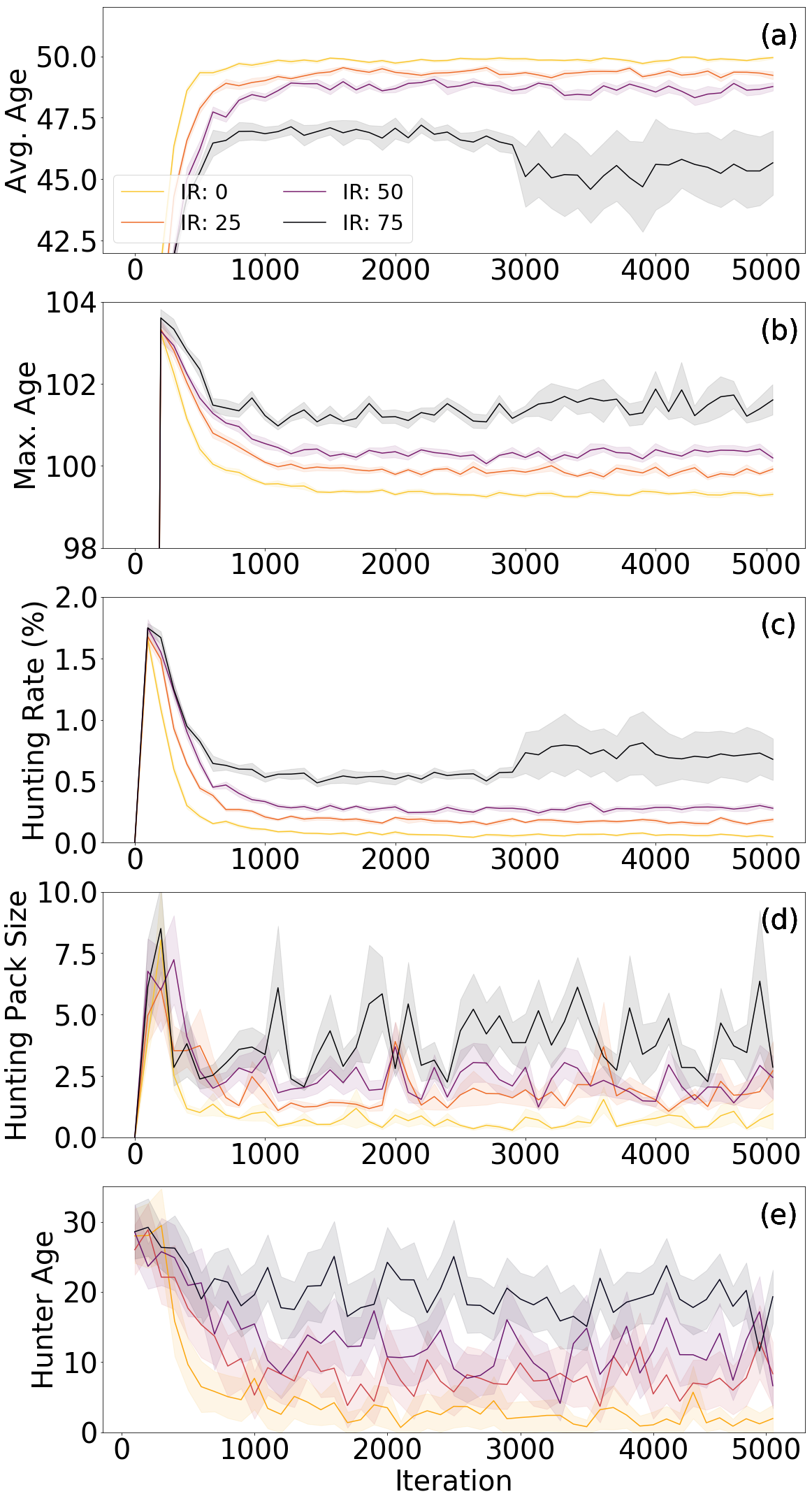}
    \caption{Dynamics of the experiment results. Panel (a) shows the average age of agents. Panel (b) shows the maximum age of agents. Panel (c) shows the percentage of agents that hunt at each iteration. Panel (d) shows the hunters' cluster size prior to hunting the opposing agent. Panel (e) shows the age of the agents hunting. Color is proportional to the interdependence reward (IR). Darker color indicates higher interdependence reward. Each plot is obtained by averaging results of ten experiments and 100 iterations. Shaded areas denote the standard deviation among the experiments. }
    \label{fig:avg_age}
\end{figure}

Experiments are conducted by setting up different values of incentives and observing the emergent collective behavior associated with each experiment. During simulations, agents explore the space of possible behaviors and inform which behaviors are promoted under certain incentives and environmental rules. As a result, we create an artificial environment for testing hypotheses and obtaining information through simulations hard to anticipate given the complexity of the space of possibilities.\footnote{Demonstration of the experiments: 
(IR: 0) \url{https://youtu.be/AgAeYMe2tUE} 
(IR: 25) \url{https://youtu.be/OZbl8qD50Mg} 
(IR: 50) \url{https://youtu.be/Ca2p2cATmlw} 
(IR: 75) \url{https://youtu.be/R32Xu_EUpBQ}. }

In this case, we create agents who want to segregate from other kinds and provide incentives to create interactions and interdependencies across kinds. For this purpose, we model the Schelling dynamics for segregation and combine it with the interdependence reward. The interdependence reward is given when agents of different kinds compete and win against each other following hunting dynamics. The one who is hunted dies and the hunter gets a positive reward and life-extension. In total, there are four different experiments with interdependence reward of 0, 25, 50 and 75 respectively. A set of videos are available with one simulation for each setting. In the videos, colors yellow/orange and cyan/magenta denote the types of agents. The color brightness indicates the age of agents for both kinds. 

Interdependence rewards diminish spatial segregation among different types. In Figure \ref{fig:heat}a we show the collective behavior of the population, using heat maps proportional to the probability of agents location during simulations according to their type. The heat maps are visualized over one trial of the experiments. Blue and red regions show biases towards each kind. White regions show uniform occupation. The dynamics of segregation quickly result in patches of segregated groups (top panels). As interdependence rewards increase, the probability of one grid being occupied by agent of type A or B becomes uniform and plots become white (bottom right panels). By creating interdependencies among agents, they increase their interactions and reduce the spatial segregation. 

We measure segregation among agents using multiscale entropy. We convolve the grid space with low pass filters of size 6x6, 12x12 and 25x25 using sliding windows whose output is the window average value. We measure the entropy of the distribution of window averages after each convolution across all iterations.  The segregation per iteration is defined as the average entropy across the distributions resulting from the different filter sizes. The resulting segregation dynamics is visualized in Figure \ref{fig:segregation}. Segregation is high when interdependencies are not rewarded (yellow curve). As interdependencies increase (purple and black curves), the agents mix and the spatial segregation is significantly reduced ($p<0.001$, see Section \ref{sup:test} in the Supplement).

Interdependencies affect the group dynamics. As we increase the reward for interdependencies, the initially stable patches emerging from promoting segregation become dynamic and mix with the other kind. The properties of the population and associated  activities reflect the change of dynamics. Agents create an internal hierarchy where younger agents go out and hunt and elder agents segregate and ensure reproduction. Evidences of such behavior are that the average age of agents decreases and the hunting rate increases (Figure \ref{fig:avg_age}a and \ref{fig:avg_age}c) and average hunter age is much lower than the average agent age (Figure \ref{fig:avg_age}a and \ref{fig:avg_age}e). Moreover, the maximum age of agents per kind increases (Figure \ref{fig:avg_age}b) showing that some agents stay protected and do not hunt. The hunting strategy of agents is also affected by increasing interdependencies. Pack size increases consistently with interdependence rewards.  Figure \ref{fig:avg_age}d shows the size of hunting clusters one step before hunting an agent. The increasing cluster size given interdependence rewards suggests that agent association yields better results. It also shows that hostile systems favor agglomeration of agents for safety which can result in ultimate polarization. Additionally, we also analyzed the effects of the vigilance rewards on the dynamics for multiple reward values. Results show that higher vigilance reward increases intra-kind interaction and results in more segregation (see Section \ref{sup:VR}).

Diverse areas attract younger people and people are older in segregated areas. We show that older agents are more segregated than younger ones in the model (see Figure \ref{fig:heat}b). The behavior has been observed in the model and verified with human behavior using Census data. We analyzed the relationship between age and segregation using Census data across the whole US (see Section \ref{sup:age}). A segregation metric based on racial entropy correlated positively with median age by census tract (r=0.4). Our simulation shed light on an observation that is not trivial about current societies. 


In summary, our experiments show that increasing interdependencies among kinds can be applied to reduce segregation. Moreover, hostile interdependencies will result in in-group cooperation for hunting and competition for sheltering. The emergent behavior of the population can be framed in the exploiter and explorer discussion. A part of it chooses to segregate and another one to go out and explore. The one who explores hunts and is vulnerable to be hunted, but creates spatial integration. The one who segregates lives longer and ensures reproduction of its own kind. In this model, explorers tend to be younger and keepers tend to live longer. Spatial mixing was achieved by increasing interaction rewards but was accompanied by larger clusters of agents of the same size. Polarization may arise when there is an adversarial relationship between the parts that segregate from each other. More generally, emergent behaviors lie in a non-linear space where interaction properties determine outcomes which may happen simultaneously and in different combinations.

\section{Discussion}
\label{discussion}

We created an artificial environment for testing rules of interactions and incentives by observing the behaviors that emerge when applied to multi-agent populations. Incentives can generate surprising behaviors because of the complexity of social systems. 
As problems become complex, evolutionary computing is necessary to achieve sustainable solutions. We combine system modeling (ABMs) with artificial intelligence (RL) in order to explore the space of solutions associated to promoted incentives. RL provides ABMs the information processing capabilities that enables the exploration of strategies that satisfy the conditions imposed by the interaction rule. In turn, ABMs provide RL with access to models of collective behavior that achieve emergence and complexity. While ABMs provide access to the complexity of the problem space, RL facilitates the exploration of the solution space. Our methodology opens a new avenue for policy makers to design and test incentives in artificial environments.

\section*{Acknowledgements}
We would like to thank Intel AI DevCloud Team for granting access to their cloud with powerful parallel processing capabilities. Also, we would like to thank Dhaval Adjodah for his valuable suggestions on training RL algorithms.

\section*{Authors Contributions}

ES, YBY and AJM contributed equally in the conceptualization, development and interpretation of the experiments as well as in the paper write up.

\section*{Data Availability}

The source code of the model implementation as well as the data generated to create this report will be made available upon publication.

\section*{Additional Information}
We declare that have no competing interests.


\bibliography{example_paper}


\newpage

\section*{Supplement}

\renewcommand{\thesection}{S\arabic{section}}
\renewcommand\thefigure{S\arabic{figure}}    
\renewcommand\thetable{S\arabic{table}}    
\setcounter{figure}{0} 
\setcounter{section}{0} 
\setcounter{table}{0} 

\section{Future Work}
\label{sup:future}

There are many potential improvements to our work. We classify directions of future work under three categories: representation, training and experimentation. Our method can be advanced by representing agents more realistically such as introducing heterogeneous personalities to agents or facilitating network structure over agents to promote alliances. Moreover, training RL agents yield better results with sophisticated exploration strategies \cite{nikolov2018information, tang2017exploration, fu2017ex2}. In addition to exploration strategies, MARL is shown to perform better with curriculum learning \cite{bansal2017emergent}. Our aim is to extend the work on multi agent curriculum learning to our problem. 

Schelling and Predator - Prey models cover just a little portion of the ABM domain \cite{macy2002factors}. We are currently working on extending this artificial environment to other ABMs, i.e. Axelrod model \cite{axelrod1997dissemination}. Our goal is to develop an easy interface where policy makers and AI researchers can collaborate on solving societal problems. 

\section{Statistical Significance}
\label{sup:test}

 We validate the significance of the patterns we observe along the execution of the simulation as we change the IR incentive in Figure \ref{fig:segregation}. We analyze the distribution of values across the last 1000 interactions for each IR values and test the difference among their averages. In Table \ref{supp:test_seg} we summarize the results of the statistical tests. The differences in averages are statistically significant ($p<0.001$) across all pairs of curves.

\section{Vigilance Reward}
\label{sup:VR}

\begin{figure}
    \centering
        \includegraphics[width=\textwidth]{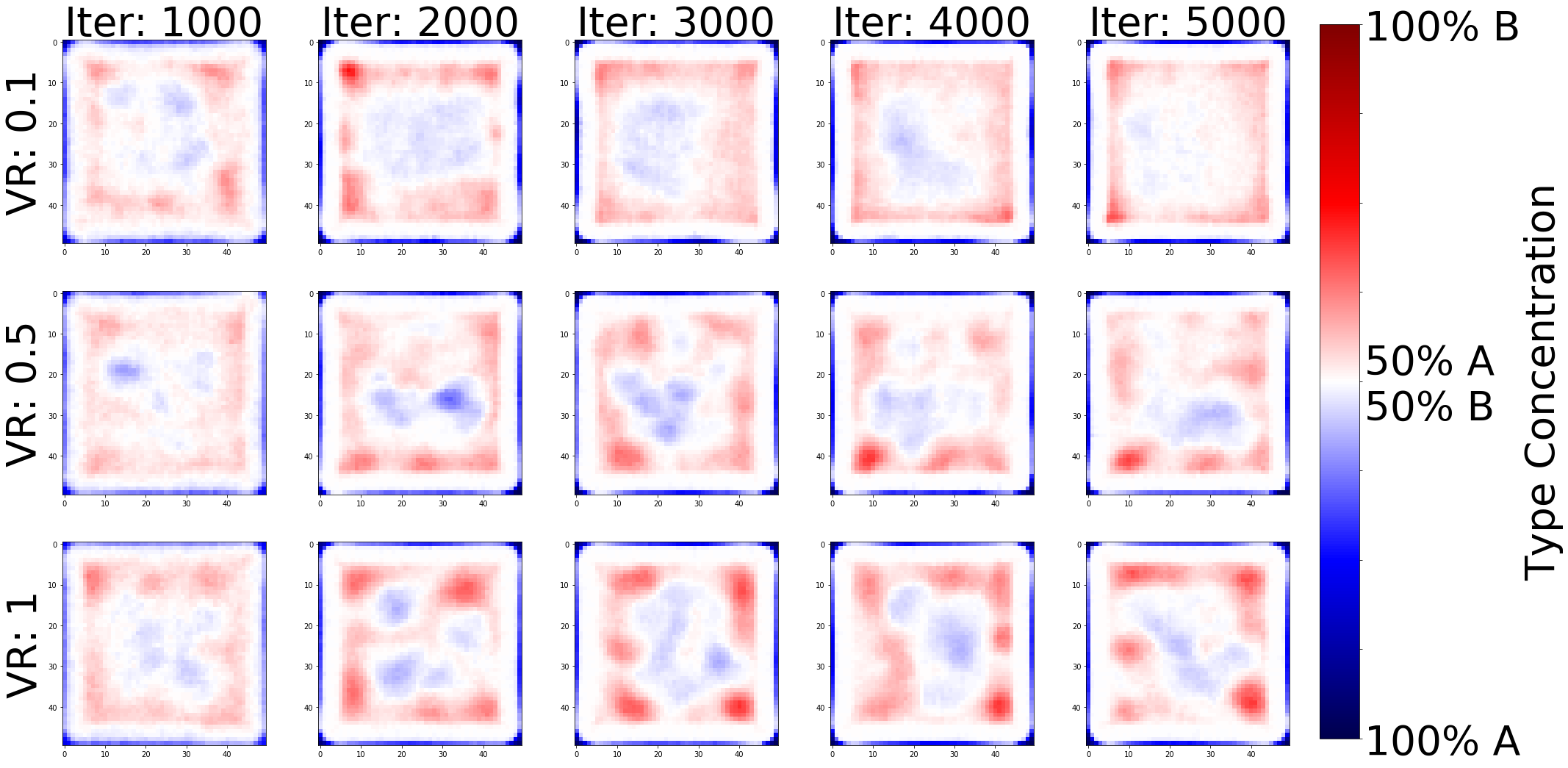}
    \caption{ Spatial distribution of agent types with varying Vigilance Reward (VR) (vertical) and Iteration (horizontal). Color indicates concentration of each type.
}
    \label{supp:VR_type}
\end{figure}

\begin{figure}
    \centering
    \includegraphics[width=\textwidth]{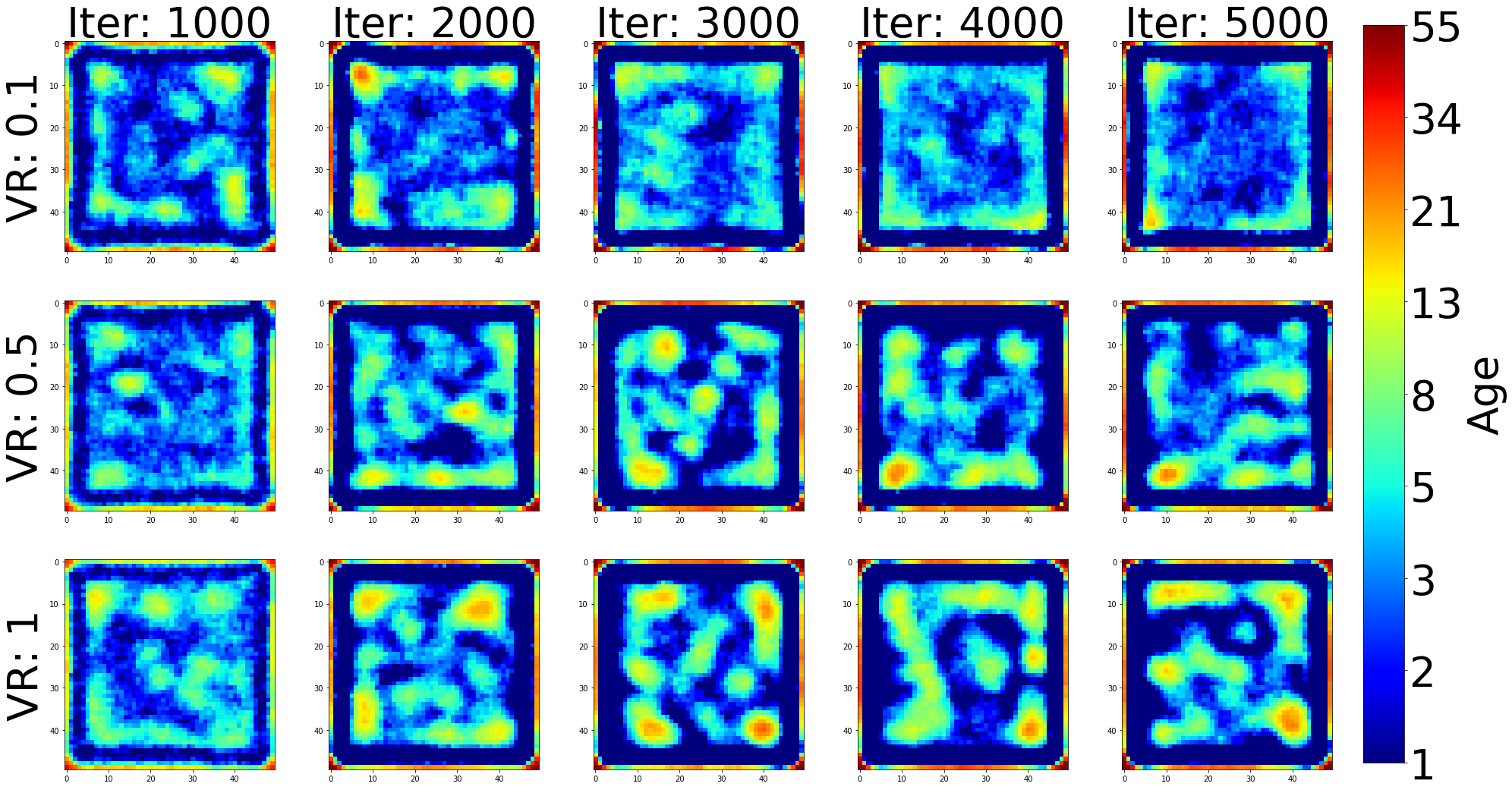}
    \caption{Spatial distribution of agent ages with varying Vigilance Reward (VR) (vertical) and Iteration (horizontal).
}
    \label{supp:VR_age}
\end{figure}

We analyze the effects of the Vigilance Rewards (VR) on the dynamics of agents. In Figures \ref{supp:VR_type} and \ref{supp:VR_age} we the impact on segregation and age distribution for multiple values of VR. The results show that increased VR increases intra-kind behavior and as a results increases segregation. Therefore, segregation may also be fostered by other types of behaviors. 

\section{Agents Age}
\label{sup:age}

\begin{figure}
    \centering
    \includegraphics[width=\textwidth]{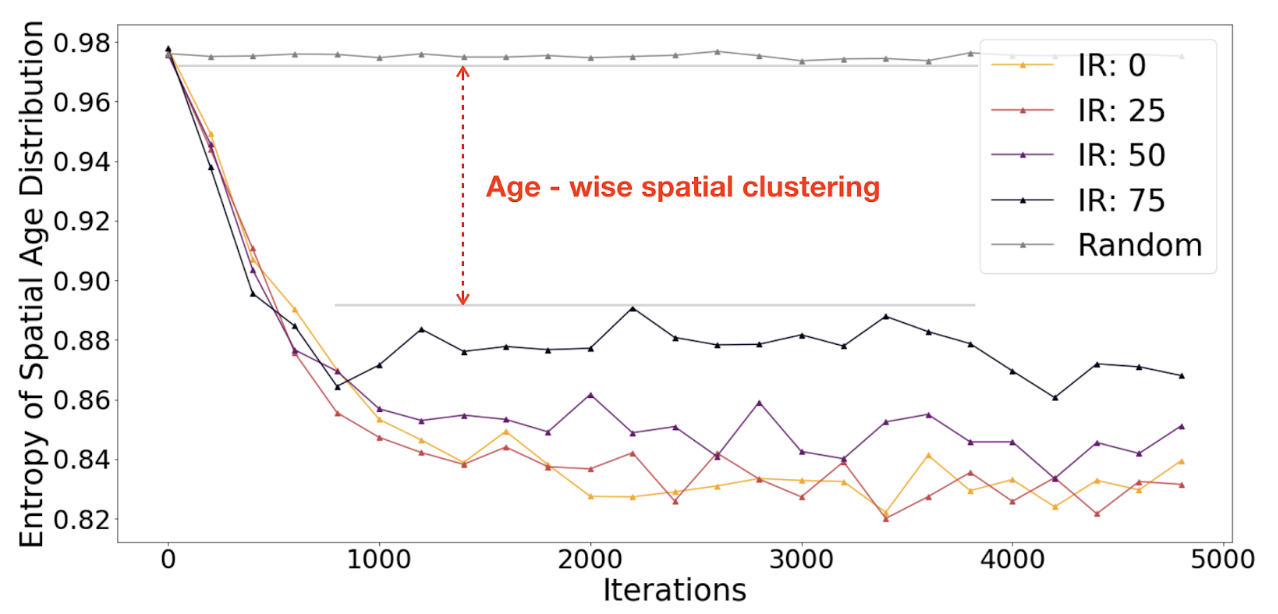}
    \caption{Entropy of Spatial distribution of agent ages with iteration. Random case is constructed by drawing agent ages on the grid from a uniform distribution and calculating the entropy. The difference in entropy corresponds to spatial clustering of ages. Entropy is normalized by the logarithm of the grid size.}
    \label{supp:age_entropy}
\end{figure}

We analyze the significance of the spatial distribution of agent ages shown in Figure \ref{fig:heat}b. In Figure \ref{supp:age_entropy} we show the entropy of the spatial distribution of agent ages at multiple iteration times and interdependence reward (IR), together with a randomized case for comparison. The randomized case is constructed by drawing agent ages on the grid from a uniform distribution between (0, 1) and calculating the entropy. The difference between the random case and the empirical results is significant in all cases. We tested significance by comparing each curve. A summary of the test results are presented in Table \ref{supp:age_entropy_test}.

\begin{figure}
    \centering
    \includegraphics[width=\textwidth]{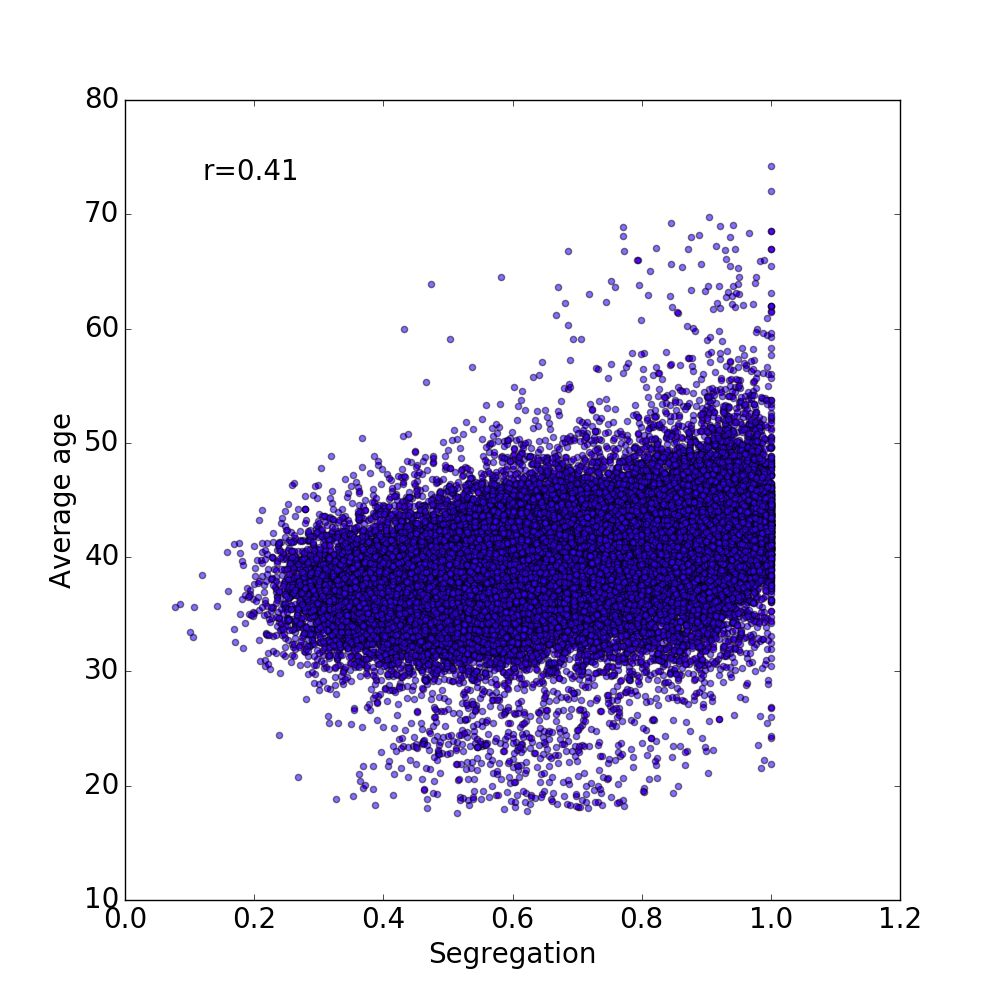}
    \caption{Age and racial segregation by census tract (dots). Pearson correlation r annotated in the Figure.}
    \label{supp:age_seg}
\end{figure}

The behavior has been observed in the model and verified it with human behavior using Census data. We analyzed the relationship between age and segregation using Census data across the whole US. A segregation metric based on racial entropy correlated positively with median age  by census tract (r=0.4). In Figure \ref{supp:age_seg} we present a scatter plot of the segregation metric (x-axis) and average age (y-axis) of each census tract (dots).

\newpage

\begin{table}[]
\centering
\begin{tabular}{|l|l|l|l|}
\hline
$R_1$ & $R_2$ & t-value & p-value  \\ \hline
0  & 25 & 4.373   & 6.44e-06 \\ \hline
0  & 50 & 15.935  & 0.0      \\ \hline
0  & 75 & 23.890  & 0.0      \\ \hline
25 & 0  & -4.373  & 6.44e-06 \\ \hline
25 & 50 & 11.267  & 0.0      \\ \hline
25 & 75 & 20.331  & 0.0      \\ \hline
50 & 0  & -15.925 & 0.0      \\ \hline
50 & 25 & -11.267 & 0.0      \\ \hline
50 & 75 & 11.691  & 0.0      \\ \hline
75 & 0  & -23.890 & 0.0      \\ \hline
75 & 25 & -20.331 & 0.0      \\ \hline
75 & 50 & -11.691 & 0.0      \\ \hline
\end{tabular}
\caption{Statistical tests results comparing segregation outcomes for different values of Interdependence Reward denoted as $R_1$ and $R_2$. We tested the difference in segregation over the last 1000 iterations for every pair of curves. The tests show that the average segregation differs across curves of different rewards.}
\label{supp:test_seg}
\end{table}

\begin{table}[]
\centering
\begin{tabular}{|l|l|l|l|}
\hline
$V_1$     & $V_2$     & t-value & p-value \\ \hline
Random & IR: 75 & 18.755  & 0.0     \\ \hline
Random & IR: 50 & 17.105  & 0.0     \\ \hline
Random & IR: 25 & 16.347  & 0.0     \\ \hline
Random & IR: 0  & 15.665  & 0.0     \\ \hline
IR: 75 & Random & -18.755 & 0.0     \\ \hline
IR: 75 & IR: 50 & 2.676   & 0.010   \\ \hline
IR: 75 & IR: 25 & 3.772   & 0.0     \\ \hline
IR: 75 & IR: 0  & 3.520   & 0.001   \\ \hline
IR: 50 & Random & -17.105 & 0.0     \\ \hline
IR: 50 & IR: 75 & -2.676  & 0.010   \\ \hline
IR: 50 & IR: 25 & 1.214   & 0.231   \\ \hline
IR: 50 & IR: 0  & 1.039   & 0.304   \\ \hline
IR: 25 & Random & -16.347 & 0.0     \\ \hline
IR: 25 & IR: 75 & -3.772  & 0.0     \\ \hline
IR: 25 & IR: 50 & -1.214  & 0.231   \\ \hline
IR: 25 & IR: 0  & -0.144  & 0.886   \\ \hline
IR: 0  & Random & -15.665 & 0.0     \\ \hline
IR: 0  & IR: 75 & -3.521  & 0.001   \\ \hline
IR: 0  & IR: 50 & -1.039  & 0.304   \\ \hline
IR: 0  & IR: 25 & 0.144   & 0.886   \\ \hline
\end{tabular}
\caption{Statistical significance of comparing the average age entropy for different Interdependence Reward, denoted by $V_1$ and $V_2$. Random curves are denoted by sampling entropy value per iteration from a uniform distribution between zero and one.}
\label{supp:age_entropy_test}
\end{table}

\end{document}